\documentclass[aps,amsmath,amssymb,superscriptaddress,floatfix]{revtex4}
\usepackage{amsmath,amssymb,amsfonts,mathrsfs}
\usepackage{epsf}
\usepackage{psfrag}
\usepackage{graphicx}
\usepackage{color}
\newcommand\beq{\begin{equation}}
\newcommand\eeq{\end{equation}}
\newcommand\bea{\begin{eqnarray}}
\newcommand\eea{\end{eqnarray}}

\newcommand{\non}{\nonumber}
\newcommand{\bib}{\bibitem}
\newcommand{\al}{\alpha}

\newcommand{\de}{\delta}
\newcommand{\De}{\Delta}
\newcommand{\ep}{\epsilon}
\newcommand{\om}{\omega}

\newcommand{\si}{\sigma}
\newcommand{\da}{\dagger}

\newcommand{\la}{\langle}
\newcommand{\ra}{\rangle}

\begin{document}

\title{Topological phases, Majorana modes and quench dynamics in a spin ladder
system}
\author{Wade DeGottardi}
\affiliation{Department of Physics, University of Illinois at
Urbana-Champaign, 1110 W.\ Green St.\ , Urbana, IL  61801-3080, USA}
\author{Diptiman Sen}
\affiliation{Centre for High Energy Physics, Indian Institute of
Science, Bangalore 560012, India}
\author{Smitha Vishveshwara}
\affiliation{Department of Physics, University of Illinois at
Urbana-Champaign, 1110 W.\ Green St.\ , Urbana, IL  61801-3080, USA}

\begin{abstract}
We explore the salient features of the `Kitaev ladder', a two-legged
ladder version of the spin-$1/2$ Kitaev model on a honeycomb lattice,
by mapping it to a one-dimensional fermionic $p$-wave superconducting
system. We examine the connections between spin phases and
topologically non-trivial phases of non-interacting fermionic
systems, demonstrating the equivalence between the spontaneous
breaking of global $\mathbb{Z}_2$ symmetry in spin systems and the
existence of isolated Majorana modes. In the Kitaev ladder, we
investigate topological properties of the system in different sectors
characterized by the presence or absence of a vortex in each
plaquette of the ladder. We show that vortex patterns can yield a
rich parameter space for tuning into topologically non-trivial
phases. We introduce a new topological invariant which explicitly
determines the presence of zero energy Majorana modes at the
boundaries of such phases. Finally, we discuss dynamic quenching
between topologically non-trivial phases in the Kitaev ladder, and in
particular, the post-quench dynamics governed by tuning through a
quantum critical point.
\end{abstract}
\pacs{75.10.Jm,03.65.Vf}

\maketitle

\date{\today}

\section{Introduction}

It has been recognized recently that non-interacting fermionic
lattice systems can exhibit a rich variety of phases based on their
topological properties. On the other hand, it has been long known
that a class of spin systems can be mapped to non-interacting
fermionic systems. Here, we exploit the mapping to tailor the recent
insights in fermionic systems to the context of spin systems, in
particular, to a system which we dub the `Kitaev ladder'. The Kitaev
ladder is a quasi one-dimensional (1D) analog of the Kitaev model on
the honeycomb lattice \cite{kitaev2}, a model which has been
extensively studied for its topological aspects \cite{feng,
chenhu,baskaran,lee,kivelson,nussinov,ortiz,schmidt}; the ladder
system shares many of the important features of the parent
two-dimensional (2D) system. The ladder has two legs with spin-$1/2$
degrees of freedom at each lattice site, highly asymmetric couplings
between nearest-neighbor spins, and conserved $\mathbb{Z}_2$ degrees
of freedom on each plaquette associated with the presence or absence
of a vortex. We will show that this ladder can be mapped to the
celebrated one-dimensional $p$-wave superconducting system.

In pioneering work \cite{kitaev1}, Kitaev showed that a 1D $p$-wave
superconducting system can exhibit topologically distinct phases
characterized by a topological $\mathbb{Z}_2$ index. This index
characterizes the topology of any one-dimensional particle-hole
symmetric topological insulator~\cite{kane}. Topologically
non-trivial phases are characterized by the presence of localized
zero energy Majorana modes at the ends of an infinitely long system
having open boundary conditions; kinetic energy and superconducting
pairing, so to speak, conspire to split the Dirac fermions comprising
the system into their `real' and `imaginary' parts. Here, we discuss
the manner in which these specific topological aspects and phases
translate to the context of spin chains. Topologically non-trivial
phases in the fermionic system may be associated with symmetry-broken
phases in the spin systems, and the trivial phases with paramagnetic
phases. As observed in~\cite{feng,chenhu}, we find that the presence
of isolated Majorana modes at the ends of a long system with open
boundary conditions is intimately related to spontaneous breaking of
a global $\mathbb{Z}_2$ symmetry. We build on this observation by
explicitly demonstrating the connection between spontaneous broken
symmetry in a spin system and Majorana modes in the corresponding
fermion system. While the behavior of spin expectation values and
correlations is generally related to those of fermions in a complex,
non-local fashion, the language of topology provides an immediate and
novel perspective on spin phases.

The connection between topologically non-trivial and magnetic phases is
generic to a large class of spin systems. What distinguishes the Kitaev
ladder is the set of $\mathbb{Z}_2$ degrees of freedom associated with
different vortex sectors. In the parent Kitaev honeycomb system,
while most studies have focused on the ground state sector
corresponding to the absence of a vortex in every plaquette, studies
of other vortex sectors have been sparse. Here, the ladder system
provides a simple prototype for studying the physics of different
vortex sectors. The distribution of vortices on plaquettes
can, in the corresponding fermionic system, be encoded in the sign of
the chemical potential on each lattice site. We explore the phases
realized in a range of periodic vortex sectors and find novel
conditions for the existence of topologically non-trivial states. For
instance, unlike in the regular 1D $p$-wave superconducting system,
some sectors require a critical amount of superconductivity
(reflected by the magnitude of the superconducting gap parameter) to
host non-trivial states.

Of late, motivated by applications to topological quantum
computational schemes \cite{nayak}, there has been a surge of
interest in seeking out systems and geometries that can realize and
manipulate isolated Majorana modes
\cite{ivanov,stone,chenhu,kells,sau,sau2,oreg,alicea,shiva,kells2,hasan,neupert,akhmerov,mao,potter,chung,hosur,iosel,ganga}.
Some analogous studies in the Kitaev honeycomb system have identified
Majorana modes bound to vortices and schemes for their manipulation
\cite{kells}. In the ladder system, the isolated modes, as opposed to
being present at vortices, completely parallel the 1D $p$-wave
superconducting system in being present at the interface between
topologically trivial and non-trivial segments. Moreover, we find
that the periodic patterns mentioned above provide a new route for
finding phases and configurations that support these modes. In
principle, for translationally invariant systems, such phases can be
characterized by a $\mathbb{Z}_2$ topological index that considers
the Berry phase accumulated by the eigenvectors in traversing the
first Brillouin zone. In practice, we find that for complex periodic
patterns, such a treatment proves to be rather involved, and can be
replaced by the evaluation of a much more direct topological
invariant derived from the equations of motion.  Our method
generalizes the identification of zero energy plane waves by Wen and
Zee \cite{wen} to the case of evanescent modes.  We use this scheme
to pinpoint several different configurations of periodic patterns
that yield localized Majorana modes in the bulk or at the ends of the
Kitaev ladder. Our analysis also provides an alternate route for
isolating Majorana modes in 1D $p$-wave superconductors by applying
appropriate periodic potentials.

The manipulation of the Majorana modes requires dynamically changing
the parameters associated with the Hamiltonian describing the system.
We consider dynamics from two angles. Based on the parallel between spin
and fermionic systems, we briefly outline the first steps necessary for
realizing schemes in the Kitaev ladder that are
analogs of those recently proposed in the 1D $p$-wave superconductor. Our
second study of dynamics entails quenching between topologically trivial and
non-trivial regions, an operation that is necessary for several
dynamic schemes. The quench dynamics is interesting in its own right
from the perspective of non-equilibrium quantum critical phenomena,
and we find that it results in the generation of residual energy or
defects whose density scales as $1/\sqrt{\tau}$ in the thermodynamic
limit, where $1/\tau$ is the quench rate. This defect production, we
believe, also has implications for finite sized regions and could
contribute to quantum decoherence.

Our presentation is arranged as follows. In section
\ref{sec:fermion_spin}, we recapitulate the key features of the 1D
$p$-wave superconductor and its topological aspects. We then show its
connections to spin physics in the context of the extensively
spin-1/2 $XY$ chain subject to a transverse magnetic field. In
section \ref{sec:KL}, we introduce the model of interest, the Kitaev
ladder, map it to the 1D $p$-wave superconductor and outline
techniques for isolating Majorana modes in this system. In section
\ref{sec:vortex_Majorana}, we present various vortex sectors in the
Kitaev ladder system, the phases realized in these sectors, and the
conditions for isolating Majorana modes. We discuss dynamical aspects
in section \ref{sec:dynamics} and, in section \ref{sec:appl_1Dpwave},
the manner in which our studies and findings would translate back to
the 1D $p$-wave superconductor.

\section{Connections between 1D $p$-wave superconductors and spin chains}
\label{sec:fermion_spin}

\subsection{The 1d $p$-wave superconductor: recapitulation}
\label{sec:recap}

\begin{center}
\begin{figure}[htb]
\includegraphics[width = 15cm]{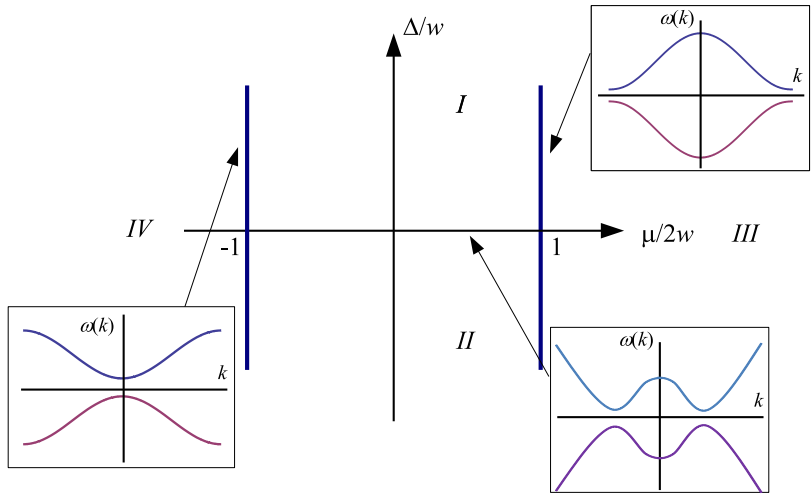}
\caption{Zero temperature phase diagram of the 1D $p$-wave
superconducting fermionic system as well as the spin-1/2 $XY$ chain
in a transverse magnetic field. Phases I and II denote topologically
non-trivial/ferromagnetic phases while phases III and IV denote
topological trivial/paramagnetic phases. Inset: Dispersion relations
near various gap closures at phase boundaries.} \label{fig:phases}
\end{figure}
\end{center}

The 1D $p$-wave superconducting system of spinless fermions explored
by Kitaev \cite{kitaev1} is described by the tight-binding
Hamiltonian \beq  H = \sum_n \left[-w \left(f_n^\da
f_{n+1}^{\phantom\da} + f_{n+1}^\da f_n^{\phantom\da} \right) + \De
\left(f_n^{\phantom\da} f_{n+1}^{\phantom\da} + f_{n+1}^\da f_n^\da \
\right) - \mu\left( f_n^\da f_n^{\phantom\da} - 1/2 \right)\right],
\label{eq:1DpwaveHam} \eeq where $w$ is the nearest-neighbor hopping
amplitude, $\De$ the superconducting gap function (assumed real), and
$\mu$ the on-site chemical potential. The translationally invariant
system can be diagonalized in the momentum basis, $f_k =
\frac{1}{\sqrt{N}} \sum_n f_n e^{-i kn}$ and shown to have the
particle-hole symmetric dispersion \beq \om_k ~=~ \pm ~\sqrt{(2 w
\cos k + \mu)^2 + 4 \De^2 \sin^2 k}. \label{eq:pwave_disp} \eeq As
shown in Fig.~\ref{fig:phases}, the system hosts distinct phases
depending on the parameter $|\mu|/2w$; the system is gapped save for
the regions demarcating the phase boundaries. The weak
(\emph{intra}-site) pairing phase $|\mu|/2w<1$ (phases I and II in
Fig.~\ref{fig:phases}) is topologically non-trivial while the strong
pairing phase (phases III and IV) is topologically trivial.

Topologically trivial versus non-trivial phases are distinguished by
the absence versus presence of boundary Majorana modes, which can be
visualized in a simple fashion in the extreme strong and weak pairing
limits as follows. Consider decomposing the Dirac $f$-fermion above
in terms of two Majorana fermions $f_n = \left( a_n + i b_n
\right)/2,$ where the Majorana fermions respect the relations $a_n^2
= b_n^2 = 1$, $\{a_n,a_m\} = \{b_n,b_m\} = 2 \de_{m,n}$, $\{a_n,b_m
\} = 0$. In terms of the Majorana fermions, the Hamiltonian in
equation (\ref{eq:1DpwaveHam}) becomes \beq H ~=~ \frac{i}{2} ~\sum_n
\left[\left(-w + \De \right) a_n b_{n+1} ~+~ \left( w + \De \right)
b_n a_{n+1} ~-~ \mu a_n b_n \right], \label{eq:hammaj} \eeq as
represented in Fig. \ref{fig:MLadder}(a).

In the extreme weak pairing limit $\De =w>0$, $\mu=0$, the
tight-binding Hamiltonian in (\ref{eq:1DpwaveHam}) couples the
Majorana fermions in the staggered fashion shown in Fig.
\ref{fig:MLadder}(c). A change in sign of $\De$ corresponds to
shifting the staggered pattern by a lattice site, thus exchanging the
roles of the $a$ and $b$ fermions. The superconductivity induced
anomalous term and the normal hopping term conspire to separate a
Dirac fermion into its Majorana components, leaving an isolated
Majorana mode at each end. In this limit, equation
(\ref{eq:1DpwaveHam}) decouples into the form
$\sum_{n=1}^{N-1}(w+\De)(\tilde{f}^{\da}_n
\tilde{f}_n^{\phantom\da}-1/2)$, where $\tilde{f}_n=(a_n+i
b_{n+1})/2$ corresponds to Dirac fermions composed of the pairs of
linked Majorana fermions depicted in Fig. \ref{fig:MLadder}(c). This
then yields two Majorana fermions $b_1$ and $a_N$ which are isolated
at the two ends and which do not appear in the Hamiltonian since
these modes have zero energy \cite{kitaev1}. In the extreme strong
pairing limit $\De=w=0$, $\mu\neq 0$, the Majorana fermions $a_n$ and
$b_n$ are pairwise connected as shown in Fig. \ref{fig:MLadder}(b);
thus no Majorana fermions are isolated.

The presence of an energy gap in each phase ensures that slightly
changing the couplings from these extreme limits does not alter these
topological aspects. In the weak pairing phase, it is thus still
possible to define appropriate linear combinations $Q_L=\sum\al_n
a_n$ and $Q_R=\sum\beta_n b_n$ (with $\al_n$ and $\beta_n$ real) that
are Majorana modes bound to the two ends of the system and that
become isolated in the thermodynamic limit.

\subsection{Mapping spin chains to a $p$-wave superconductor}
\label{sec:mapping}

\begin{figure}[htb]
\includegraphics[width = 10cm]{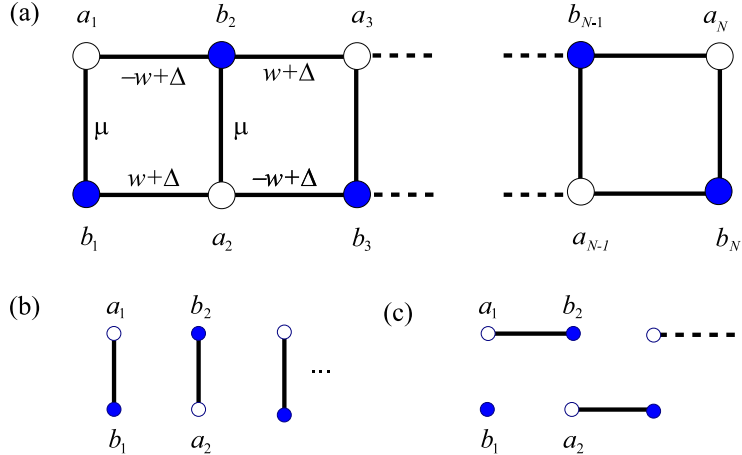}
\caption{Pictorial representation of the couplings between Majorana fermions
described by the Hamiltonian in equation (\ref{eq:hammaj}) for the cases of
(a) general couplings, (b) the strong (intra-site) pairing limit $\De = w = 0$,
$\mu \neq 0$ which does not possess Majorana modes (topologically trivial),
and (c) the weak pairing limit $\De = -w$, $\mu = 0$ which hosts Majorana
modes at the ends (topologically non-trivial).}
\label{fig:MLadder} \end{figure}

The concepts discussed above can be investigated in the context of
spin chain physics. As an example, consider the extensively studied
spin-1/2 $XY$ chain subject to a transverse magnetic field
\cite{lieb,barouch,bunder}. The Hamiltonian is given by \beq H ~=~-
~\sum_{n=1}^{N-1} ~\left( J_x \si_n^x \si_{n+1}^x+J_y \si_n^y
\si_{n+1}^y\right) ~-~ h ~\sum_{n=1}^N ~\si_n^z, \label{ham1} \eeq
where $\si_n^\al$ denote the Pauli matrices. The spin-$1/2$ operators
can be mapped to Majorana fermions using the Jordan-Wigner
transformation \cite{lieb,kogut} $a_n = \prod_{j=1}^{n-1} ~\si_j^z
~\si_n^x, b_n = \prod_{j=1}^{n-1} ~\si_j^z ~\si_n^y$, for $2 \le n
\le N$, $a_1 = \si_1^x$ and $b_1 = \si_1^y$. The resultant
Hamiltonian exactly maps to the 1D $p$-wave superconductor Majorana
Hamiltonian of equation (\ref{eq:hammaj}) with the identification $w
\leftrightarrow J_x+J_y, ~\De \leftrightarrow J_y-J_x, ~\mu
\leftrightarrow 2h$, and with the interchange of $a_n$ and $b_n$ on
even sites. Hence, Fig. \ref{fig:phases} also represents the phase
diagram for the spin system. Phases I and II are ferromagnetic and
have non-zero ground state expectation values of $\si_n^x$ and
$\si_n^y$, respectively. Phases III and IV are paramagnetic with both
$\si_n^x$ and $\si_n^y$ having zero expectation values in the ground
state. The extreme limits shown in Fig.~\ref{fig:MLadder} can be
understood in terms of spin physics. The case of
Fig.~\ref{fig:MLadder}(c) corresponds to the values of the couplings
$J_x\neq 0, J_y=0, h=0$ and thus to an Ising-type ferromagnetic (or
antiferromagnetic) ground state along the $x$-axis. The case of
Fig.~\ref{fig:MLadder}(b) corresponds to $J_x = 0, J_y=0, h\neq0$ and
thus to spins polarized along the $z$ direction.

\subsection{Spontaneous symmetry breaking and isolated Majorana modes}
\label{sec:symbreaking}

As discussed above, the topologically non-trivial phases in the
$p$-wave superconductor, associated with isolated boundary Majorana
fermions, map onto the ferromagnetic phases in the spin chain. We
argue here as follows that the presence of these isolated fermions
corresponds to a $\mathbb{Z}_2$ symmetry in the spin phases which
gets broken spontaneously. As is obvious in the extreme limit of Fig.
\ref{fig:MLadder}(c), the energetics in phases I and II provides
$N-1$ constraints for a system of $N$ spins. Hence, the ground state
is constrained to two degenerate states reflecting a $\mathbb{Z}_2$
symmetry. Returning to the language of fermions, the degeneracy is
associated with the isolated modes $Q_L=\sum\al_n a_n$ and
$Q_R=\sum\beta_n b_n$ which do not participate in the energetics.
Explicitly, the Dirac fermion formed by the two boundary Majorana
fermions, $\Psi=(Q_L+ i Q_R)/2$ can have occupation number $0$ or
$1$. Thus, picking one of the corresponding degenerate states
$|0\rangle_\Psi$ or $|1\rangle_\Psi$ (or any linear combination
$u|0\rangle_\Psi + v|1\rangle_\Psi$) amounts to breaking the
$\mathbb{Z}_2$ symmetry of the ground state and picking one of the
two energetically available spin configurations. In phases III and
IV, however, energetics poses $N$ constraints and the system is
confined to a unique ground state.

To explore the spin physics in the symmetry broken phase in terms of
the isolated Majorana modes, consider the spin-$1/2$ algebra formed
by $\{ Q_L, Q_R, - iQ_LQ_R \}$, where $i Q_LQ_R=2\Psi^{\da}\Psi-1$.
The two eigenstates of any of these operators (those of the $-i
Q_LQ_R$ being $|0\rangle_\Psi$ and $|1\rangle_\Psi$) are orthogonal
to one another and are each equally valid configurations in the
ground state. The particular symmetry broken choice immediately
determines the expectation values of these operators and their
corresponding Majorana and spin operators. Specifically, if we
consider the two eigenstates of $Q_L$, $|\psi_{\pm} \ra_L$, we can
use the relationship $\{ Q, a_n \} = 2 \al_n$ to show that $\la
\psi_+| a_n |\psi_+ \ra_L = \al_n$ and $\la \psi_-| a_n |\psi_- \ra_L
= - \al_n$ for all $n$. Furthermore, since $\{ Q, b_n \} = 0$, we see
that $\la \psi_{\pm}| b_n |\psi_{\pm} \ra_L = 0$ for all $n$. The
physics behind these statements can be made more transparent, for
instance, in the limiting case $J_x
>|J_y|$ and $h=0$. Here, the system is in the Ising-type phase whose
possible degenerate ground states correspond to spins that are either
primarily aligned or anti-aligned along the $\hat{x}$ direction. In
fact, $|\psi_{\pm} \ra_L $ are exactly these two states. Given that
$\si_1^x$ is identically $a_1$, we can easily evaluate its
expectation value in one of the states to get $\la \psi_+|\si_1^x |
\psi_+ \ra_L ~=~ \sqrt{1-(J_y/J_x)^2}$, as ought to be true for the
system pointing primarily along $\hat{x}$.

The non-local nature of the Majorana modes is consistent with the
spontaneously broken symmetry being global. Energetics locks all
spins into one of two possible configurations where the Jordan-Wigner
string connects the chain in a non-local, highly entangled fashion. A
particular choice of states in the basis of the isolated Majorana
modes thus corresponds to picking one of these highly entangled
states. It should be noted that a knowledge of which one of these
states is chosen, for instance $|\psi_+ \ra_L$, gives limited
information in that expectation values can be evaluated only for
non-local spin operators corresponding to Majorana modes $a_n$ and
$b_n$, except for one or two operators, such as $\si_1^x$ above. This
knowledge does provide evidence for long-range order. For example,
our result for $\la \psi_+|\si_1^x | \psi_+ \ra_L$ in the example
above can be compared to that of Lieb, Schultz and Mattis
\cite{lieb}. It is shown in Ref. \cite{lieb} that, for an open chain
with $N$ sites (where $N$ is even and $\to \infty$) in a case which
is effectively that of the Ising limit above, the end-to-end
two-point correlation functions are given by $\la \psi_+| \si_1^x
\si_N^x |\psi_+ \ra = 1-(J_y/J_x)^2$ and $\la \psi_+| \si_1^y \si_N^y
|\psi_+ \ra = 0$. In the limit $N \to \infty$, we expect clustering
to hold, so that $| \la \psi_+| \si_1^x | \psi_+ \ra | = | \la
\psi_+| \si_1^x \si_N^x |\psi_+ \ra |^{1/2}$. We therefore find
complete agreement with the results in~\cite{lieb}. We note that our
derivation is much simpler than the one in Ref. \cite{lieb}, purely
making use of the end Majorana modes, and our results can be easily
generalized to take into account the transverse magnetic field $h$ in
Eq. (\ref{ham1}) which was not considered in~\cite{lieb}. Thus,
analysis of the isolated Majorana modes can be a valuable means for
gaining insight into spin chain physics.

\section{The Kitaev ladder system}
\label{sec:KL}

\subsection{The model}
\label{sec:model}

\begin{figure}[htb]
\includegraphics[width=10cm]{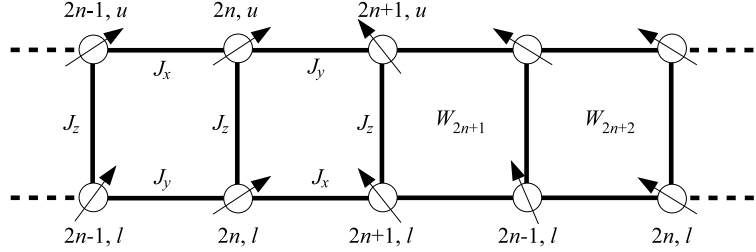}
\caption{The Kitaev ladder, the associated couplings $J_x$, $J_y$ and
$J_z$, and the vortex operators, $W_n$.}
\label{fig:spinladder} \end{figure}

Equipped with the connection between a 1D $p$-wave superconductor and
spin chains, we turn to the object of our attention, the Kitaev
ladder system. This system consists of a single (modified) strip of
Kitaev's original two-dimensional honeycomb system. The model is thus
ladder-like and each plaquette is a square instead of a hexagon. As
shown in Fig. \ref{fig:spinladder}, the two-legged ladder hosts
spin-1/2 degrees of freedom at sites $(n,u)$ and $(n,l)$, where $u$,
$l$ denote the upper and lower legs, respectively. The Hamiltonian is
given by \bea H ~=~ \sum_{n=1}^{(N-1)/2} ~[J_x \left( \si_{2n-1,u}^x
\si_{2n,u}^x + \si_{2n,l}^x \si_{2n+1,l}^x \right) \non \\
+ ~J_y \left( \si_{2n,u}^y \si_{2n+1,u}^y + \si_{2n-1,l}^y
\si_{2n,l}^y \right)] ~+~ \sum_{n = 1}^N J_z \si_{n,u}^z \si_{n,l}^z,
\label{eq:ladder} \eea where $\si^i_{n,u/l}$ denote the Pauli
matrices respecting the usual commutation rules, $\left[ \si^i_{n,s}
\si^j_{m,s'} \right] = 2 i \de_{m,n} \de_{s,s'} \ep_{i j k}
\si^k_{n,s}$.

As in the 2D honeycomb lattice system, each plaquette has an
associated operator $W_n$ which commutes with the Hamiltonian and is
of the form $\si_{n,u}^x \si_{n+1,u}^x \si_{n+1,l}^y \si_{n+1,l}^y$
for $n$ even and $\si_{n,u}^y \si_{n+1,u}^y \si_{n,l}^x
\si_{n+1,l}^x$ for $n$ odd. Since $W_n^2 = 1$, the eigenvalues of
$W_n$ are $\pm 1$; hence these invariants provide a set of
$\mathbb{Z}_2$ quantum numbers characterizing different sectors of
the Hamiltonian. These operators, commonly referred to as vortex
operators, reflect the absence/presence ($W_n=\pm 1$) of a vortex in
each plaquette. The richness of this system, its phases and
its topological properties derive from the extensive space of the
$\mathbb{Z}_2$ configurations.

\subsection{Jordan-Wigner transformation}
\label{sec:jordan}

The Kitaev ladder/honeycomb system is endowed with a special set of
couplings which enable the Jordan-Wigner transformation to provide a
{\it local} fermionic Hamiltonian \cite{feng,nussinov}. As in the
previous section, this transformation gives a direct connection
between the Kitaev ladder and the 1D $p$-wave superconductor. The
mapping relates spin operators to Majorana fermionic operators: \bea
a_n = S_{n,l/u} \si_{n,l/u}^{x/y}, & ~b_n = \pm S_{n,u/l}
\si_{n,u/l}^{x/y}, & ~\mbox{for $n$ odd/even,} \\ \non c_n =
S_{n,l/u} \si_{n,l/u}^{y/x}, & ~d_n = \pm S_{n,u/l}
\si_{n,u/l}^{y/x}, & ~\mbox{for $n$ odd/even.} \label{eq:jwKL} \eea
Here, assuming that each leg contains $N$ sites with open (rather
than periodic) boundary conditions, the string operator $S_{n,u}
=\prod_{m=1}^{n-1} ~\si^z_{m,u}$ runs from the left to the right
along the top leg in Fig. \ref{fig:spinladder} to the $(n-1)$-th
site, while $S_{n,l} = \prod_{m=1}^N ~ \si^z_{m,u} ~\prod_{m=n+1}^N
~\si^z_{m,l}$ runs completely along the top leg left to right and
then back along the bottom leg to the $(n+1)$-th site.

In terms of these Majorana operators, the Hamiltonian is of the form
$H = \sum_n \left( -i J_x a_n b_{n+1} + i J_y b_n a_{n+1} - J_z a_n
b_n c_n d_n \right)$. The role of the vortices is encoded in the
$J_z$ term. This can be seen by noting that $W_n = -( i c_n d_n) (i
c_{n+1} d_{n+1} )$ and defining $s_n = i c_n d_n = \prod_{m = n}^N
W_m c_N d_N$. Then, the Hamiltonian, as depicted pictorially in Fig.
\ref{fig:spinladder}, takes the form \beq H = i \sum_n \left( - J_x
a_n b_{n+1} + J_y b_n a_{n+1} + J_z s_n a_n b_n \right),
\label{eq:majoranaladder} \eeq where $s_n = \pm 1$ depending on the
configuration of vortices, namely $W_n=-s_n s_{n+1}$. Now, to make a
connection with the $p$-wave system, one can once more define Dirac
fermions $f_n=(a_n+ i b_n)/2$ as in the $XY$ spin chain. The
vorticity can be captured by another set of Dirac fermions, $g_n =
\left( c_n + i d_n \right)/2$; $s_n$, which we coin `site-polarity',
reflects the occupancy of these fermions via $s_n = 2g_n^\da
g_n^{\phantom\da}-1$ and determines the sign of the chemical
potential for the $f$-fermions

In terms of the Dirac fermions, the Kitaev ladder also maps onto the
1D $p$-wave superconductor form, \bea H = &-& \sum_n [ w
\left(f_n^\da f_{n+1}^{\phantom\da}
+ f_{n+1}^\da f_n^{\phantom\da} \right) \non \\
& & ~~~~~-~ \De \left(f_n^{\phantom\da} f_{n+1}^{\phantom\da} +
f_{n+1}^\da f_n^\da \ \right) ~-~ \mu_n \left(f_n^\da
f_n^{\phantom\da} - 1/2 \right)], \label{eq:diracladder} \eea
where the identification
\beq w \leftrightarrow J_x+J_y, ~~\De \leftrightarrow J_y-J_x, ~~~{\rm
and}~~~ \mu_n \leftrightarrow 2 s_n J_z \label{eq:ident} \eeq
has been made. Thus, the $f$-fermions, as in the $XY$ spin chain,
participate in the dynamics, while the $g$-fermions encode the vortex
configurations. Each set of fermions spans a Hilbert space of size
$2^N$, together comprising the Hilbert space of size $2^{2N}$ thus
accounting for all the degrees of freedom of the original $2N$ spins
of the ladder system.

Compared to the $XY$ spin chain of the previous section, one main
difference in the mapping to the $p$-wave superconductor is that the
sign of the chemical potential is position dependent. Here, we
restrict our studies to periodic patterns. For $s_n$ with period $p$,
we can employ the decomposition \beq s_n ~=~ \sum_{q=0}^{p-1} \left(
s_q e^{i 2 \pi qn/p} ~+~ s_q^\ast e^{-i 2 \pi qn/p} \right).
\label{eq:dft} \eeq In the momentum basis ($f_k = \frac{1}{\sqrt{N}}
\sum_n f_n e^{- i kn}$), the Dirac Hamiltonian describing the Kitaev
ladder then takes the form (up to an overall constant) \bea H &=&
\sum_{0 \leq k < \pi} \left[ - 2 w \cos k \left( f_k^\da
f_k^{\phantom\da} + f_{-k}^\da f_{-k}^{\phantom\da} \right) + \De
\left( e^{i k} f_k^\da f_{-k}^\da + e^{-i k} f_{-k}^{\phantom\da}
f_k^{\phantom\da} \right) \right] \non \\
& & - 2 J_z \sum_{0 \leq k < 2\pi} \sum_{q = 0}^{p-1} \left( s_q f_{k +
2 \pi q / p}^\da f_k^{\phantom\da} + s_q^\ast f_{k - 2\pi q / p}^\da
f_k^{\phantom\da} \right). \label{eq:kham} \eea

The other difference between the ladder and the $XY$ spin chain is in
the degrees of freedom. Here, the ladder Hamiltonian provides $N-1$
constraints on the energetics for a ladder of $2N$ spins. An
additional $N-1$ constraints are given by the configuration of
vortices (note that another $\mathbb{Z}_2$ symmetry is present
because the vortex eigenvalues of $W_n$ are invariant under a global
sign change of the $s_n$'s). These constraints together connect all
$2N$ spins. Hence, once again the topologically non-trivial phases, namely
those bearing isolated boundary Majorana modes, can be associated with
spontaneous symmetry breaking in the spin language; a detailed study
of the spin physics, as initiated for the $XY$ spin chain, is in order.

\subsection{Techniques for identifying isolated Majorana modes}

The fermionic Hamiltonian in Eq. (\ref{eq:diracladder}) provides a
starting point for exploring various conditions under which bulk or
boundary Majorana modes may appear. In this section, we develop a
formalism which enables us to identify these modes in a range of
vortex sectors. We avail ourselves of the transfer matrix method
which has been used extensively in 1D systems \cite{ostlund,sen} and
is well suited to study bound states. The presence of bound Majorana
states is governed by the growth or decay of the eigenfunctions of
the transfer matrix. As a way of discerning the existence of these
states, we employ an invariant which is constructed by analytically
continuing plane wave states.

\subsubsection{Transfer matrix approach}
\label{sec:eom}

The transfer matrix can be constructed by employing the Heisenberg
representation and deriving the equations of motion from the
Hamiltonian of Eq. (\ref{eq:majoranaladder}) for the time dependent
Majorana modes $a_n = \al_n e^{-i \om t}$ and $b_n = \beta_n e^{-i
\om t}$. These equations are of the form \bea \left( w - \De \right)
\al_{n-1} + \left( w + \De \right)\al_{n+1} -
\mu_n \al_n &=& - i \om \beta_n, \non \\
- \left( w + \De \right) \beta_{n-1} - \left( w - \De \right)
\beta_{n+1} + \mu_n \beta_n &=& - i \om \al_n, \label{eq:eom1} \eea
where we have invoked the identification in Eq.~(\ref{eq:ident}).

To identify Majorana modes, we focus on $\om =0$ corresponding to real
values of $\al_n$ and $\beta_n$. The equations for $\al_n$ and $\beta_n$
decouple, and they can be written in the transfer matrix form
\beq \label{eq:matrixA}
\left( \begin{array}{c}
\al_{n+1} \\
\al_n \end{array} \right) ~=~ A_n \left( \begin{array}{c}
\al_n \\
\al_{n-1} \end{array} \right), ~~\mbox{where} ~~A_n ~=~ \left(
\begin{array}{cc}
\frac{\mu_n}{\De + w} & \frac{\De - w}{\De + w} \\
1 & 0 \end{array} \right).\eeq
Similar expressions hold for the $\beta_n$'s since the respective
transfer matrices are related by $B_n=A_n^{-1}$. Knowing the behavior
of the $a$-modes thus completely determines that of the $b$-modes.

This transfer matrix formulation enables us to study the growth
versus decay of modes at the boundary of a finite piece of the ladder
or at the interface between two parts of the ladder in different
phases. In particular, for a homogeneous system having all couplings
and $s_n$ constant, this behavior is determined by the eigenvalues of
any $A_n$, while for a region having periodicity $p$ in the $s_n$'s,
it is determined by the eigenvalues of the matrix \beq {\cal A}_P ~=~
A_p ~A_{p-1}~\cdots ~A_2 ~A_1. \label{eq:AP} \eeq

Majorana modes bound to the ends of a finite chain require that both
eigenvalues of ${\cal A}_P$ be either smaller or greater than unity
in magnitude. The former case corresponds to an $a$-mode localized at
the left end and a $b$-mode at the right end. The conditions on the
eigenvalues ensure that the localized modes can simultaneously
respect constraints at the boundary and normalizability. For
instance, if the chain goes from $n=1$ to $\infty$, the condition to
be imposed on the Majorana mode at the left end is that $\al_0 =0$
and $\al_1 =c$, where $c$ is a non-zero constant to be fixed by the
normalization. The corresponding vector $\left( \begin{array}{c}
\al_1 \\
\al_0 \end{array} \right) = \left( \begin{array}{c}
c \\
0 \end{array} \right)$ can be written as a linear superposition of
the two eigenvectors of ${\cal A}_P$. We then see from
Eq.~(\ref{eq:matrixA}) that $\al_n \to 0$ as $n \to \infty$ if both
the eigenvalues of ${\cal A}_P$ are smaller than unity.

Bound Majorana states in the bulk of the chain can be realized by
juxtaposing two regions corresponding to different phases. The
highlighting feature of this system is its ability to dial in to
different phases by changing the vortex patterns. We thus consider
the interface of two semi-infinite patterns $P$ ($n<0$) and $Q$ ($n
\geq 0$) having periodicities $p$ and $q$, respectively, but with the
same values of the couplings $w$, $\Delta$, and $\mu$. Considerations
similar to those above enable us to identify two situations in which
there exists exactly one normalizable Majorana mode for $n$ tending
to both $\infty$ and $- \infty$. These correspond to (i)~${\cal A}_P$
having both eigenvalues larger than 1 in magnitude or both smaller
than 1 and ${\cal A}_Q$ having exactly one eigenvalue less than 1 in
magnitude, or (ii)~${\cal A}_P$ having exactly one eigenvalue larger
than 1 in magnitude and ${\cal A}_Q$ having both eigenvalues less
than 1 or both smaller than 1 in magnitude. Other situations yield
either no bound mode or a pair of concurrent Majorana modes.

\subsubsection{Topological invariant}
\label{sec:TI}

The relevant features of the transfer matrix structure can be gleaned
by a simple consideration of its analytic properties. This analysis
extends the application of the topological invariant introduced by
Wen and Zee \cite{wen} to the study of evanescent modes. Our
construction of the invariant is derived from $f(z)$, the
characteristic polynomial of ${\cal A}_P$. Given that ${\cal A}_P$ is
a real $2 \times 2$ matrix, a single root inside the unit disk must
lie on the real axis. Thus the index \beq \nu ~=~ - ~\mbox{sgn}
\left( f (1) f (-1) \right), ~~{\rm where}~~ f(z) = \mbox{det} ({\cal
A}_P-I z), \label{eq:topinv} \eeq is equal to $1$ if and only if
${\cal A}_P$ has exactly one eigenvalue with a magnitude less than 1.
This case of $\nu = 1$ indicates a topologically trivial state, i.e.,
one with no end Majorana modes. However, $\nu = -1$ indicates the
existence of end Majorana modes reflected by $\mathcal{A}_P$ having
both eigenvalues less than or both greater than 1 in magnitude. The
marginal case $\nu = 0$ would imply a closure of the bulk gap. In
fact, $\nu$ can only change sign if a bulk gap closes.

The invariant $\nu$ is topological in that it is derived by invoking
the continuity of $f(z)$ to infer its behavior inside the unit disk
based on its values on the unit circle. We believe that this index,
derived using physically transparent methods, can be reduced to
Kitaev's invariant based on the Pfaffian when generalized to our
ladder model. The invariant $\nu$ can be related to that of Wen and
Zee \cite{wen} by noting the form $\nu = (-1)^{n_f+1}$, where $n_f$
is the number of zeros of $f(z)$ for which $|z| < 1$; in
Ref.~\cite{wen}, the equivalence $n_f = \frac{1}{2 \pi i} \oint_{|z|
= 1} ~dz~ f'(z)/f(z)$ was used to identify plane wave zero modes.
Furthermore, $f(z)$ was replaced by the dispersion which can also be
done in our case but makes the analysis more involved. Finally, the
invariant $\nu$ can be related to the topological invariant
of~\cite{kane} in that they both locate zero modes but the former
bypasses the need to identify the eigenvectors of Eq.~(\ref{eq:kham})
and the Berry phase across the Brillouin zone.

\section{Vortex sectors and isolated Majorana modes}
\label{sec:vortex_Majorana}

We employ the transfer matrix approach and the invariant $\nu$ to
show that a diverse set of patterns in the site-polarities $s_n$, or
equivalently, vortex sectors, can host isolated Majorana modes. We
focus on three classes of vortex sectors (i) the full vortex sector
defined by $s_n=1$, (ii) the zero vortex sector defined by $s_n=(-1)^n$,
and which is associated with the ground state sector in the parent 2D
model, and (iii) patterns of $s_n$ having higher periodicity.

\subsection{Full vortex sector: $s_n=1$}

The full vortex sector maps directly to the $XY$ spin chain in the
transverse field as well as the homogeneous 1D $p$-wave
superconductor. Isolated Majorana modes have been well-studied in the
latter case and are expected to exist at the ends of wires in the
topologically non-trivial phases I and II in Fig.~\ref{fig:phases}
and at the boundary between a topologically non-trivial phase and
topologically trivial phase in the phase diagram of
Fig.~\ref{fig:phases}. We expand on these arguments to illustrate the
techniques developed here and to provide a comprehensive description
of the Kitaev ladder. In the next subsection, we also show that this
sector corresponds to the ground state sector in a certain parameter
range.

\subsubsection{End modes}
\label{endmodes}

\begin{figure}[htb]
\includegraphics[width=11cm]{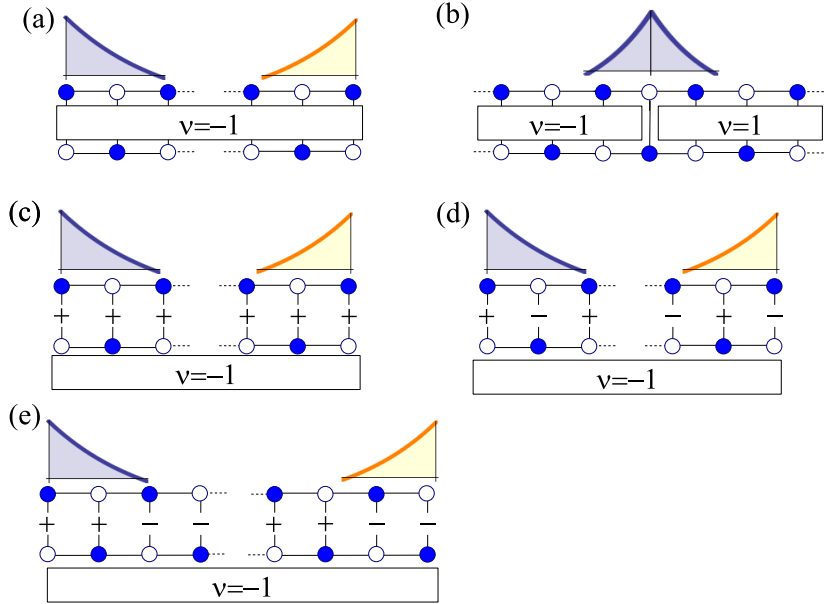}
\caption{Examples of isolated Majorana modes for the Kitaev ladder system.
Generically, Majorana modes are hosted at (a) the ends of ladders with
nontrivial topology ($\nu = -1$) or at (b) the interface between regions
with $\nu = -1$ and $\nu = 1$. In the case of the uniform sector $s_n = 1$
with $2 | w |> |\mu|$, individual Majorana modes are hosted at (c) each end
of a ladder. Other examples of end Majorana modes include
(d) the alternating sector $s_n = (-1)^n$ with $2 | \Delta | > | \mu |$ and
(e) the sector $P = + + - -$ with $ 2 \sqrt{ | w \Delta |} > | \mu |$.}
\label{fig:sectors} \end{figure}

The condition for Majorana modes to exist at the ends of the Kitaev
ladder is given by $\nu=-1$, where $\nu$ is the topological invariant of
equation (\ref{eq:topinv}). For the full vortex sector, we have a uniform
site-polarity, $s_n=1$ for all $n$; the corresponding site-independent
transfer matrix is given by
\beq {\cal A} = \left( \begin{array}{cc}
\frac{\mu}{\De + w} & \frac{\De - w}{\De + w} \\
1 & 0 \end{array} \right), \eeq which yields \beq f(z) ~=~ \mbox{det}
({\cal A} - I z) ~=~ z^2 - \left( \frac{\mu}{\De + w} \right) z ~-~
\left( \frac{\De - w}{\De + w} \right). \label{eq:fzfull} \eeq Hence,
\beq \nu ~=~ - ~\mbox{sgn} \left( f(1) f(-1) \right) ~=~ \mbox{sgn}
\left( \left| \mu \right| - 2 \left| w \right| \right),
\label{eq:condfull} \eeq for $\De + w \neq 0$. A consideration of the
location of the zeros of $f(z)$ shows that an $a$ ($b$) Majorana mode
is localized to the right (left) side of the system if $2w > |\mu|$
and $\De > 0$. If $2w < -|\mu|$ the positions of the $a$ and $b$
Majoranas are switched. Switching the sign of $\De$ also interchanges
the Majorana modes (see Fig.~\ref{fig:sectors}).

\subsubsection{Interface modes}

As an illustrative example, consider the bound modes obtained by
juxtaposing two regions in phases I and III on the left and right,
respectively, in the phase diagram of Fig. \ref{fig:phases}. The
bound state structure at the interface can be derived by considering
the closing of the gap at the phase boundary. Close to this region,
the dispersion in equation (\ref{eq:pwave_disp}) takes the form
$\om_k = \pm 2 \sqrt{ \De^2 (\de k)^2 + m^2}$ at $k \approx \pi$,
where $m(x) = \mu/2 - w$. Here, the continuum version of the
equations of motion, equation (\ref{eq:eom1}), is given by
$\partial_x a(x) = m(x) a(x)$ and $\partial_x b(x) = - m(x) b(x)$. In
this case, the Majorana mode that satisfies normalizability takes the
form $b(x) \propto e^{-\int_0^x dx' m(x')}$. In fact, this solution
corresponds to that of the celebrated Jackiw and Rebbi model
\cite{jackiw} as applied to Majorana fermions. The instances of a
single bound Majorana mode at an interface are exhausted by the cases
involving the juxtaposition of phases I/II ($|\mu|<2|w|$) and phases
III/IV ($|\mu|>2|w|$). In terms of the invariant form in
Eq.~(\ref{eq:fzfull}), these cases correspond to juxtaposing a
topologically non-trivial ($\nu=-1$) phase with a topologically
trivial phase ($\nu=1$).

\subsection{Vortex-free sector: $s_n=(-1)^n$}
\label{alternating}

In the 2D honeycomb system, the vortex-free sector is special in that it
corresponds to the ground state sector for all values of spin couplings.
This condition, however, holds only for a range of parameters in the Kitaev
ladder. To elucidate this, consider the limit of $J_z$ large compared to
$J_x$ and $J_y$. In the language of a $p$-wave superconductor, let us assume
that $\mu$ is large and positive. We then find that the perturbative change
in the ground state energy compared to the $w=\De=0$ limit is given by the
second order expression
\beq \De E_0 ~=~ - ~\frac{1}{2\mu} ~\sum_n ~\left[ w^2 (1-s_n s_{n+1}) ~+~
\De^2 (1+s_n s_{n+1}) \right]. \eeq
If $w^2 > \De^2$, the ground state energy is lowest if $s_n
s_{n+1} = -1$ for all $n$, i.e., if $s_n = (-1)^n$; this is the
sector with no vortices. If $w^2 < \De^2$, however, the ground state energy
is lowest if $s_n s_{n+1} = 1$ for all $n$, i.e., if $s_n = \pm 1$ for all
$n$; this is the full vortex sector. Another difference is that
unlike in the 2D system, reduced dimensionality renders the Kitaev ladder
system to be gapped everywhere except along phase boundaries.

For the vortex free sector, we can identify the range of parameter
space which yields topologically non-trivial phases. Towards this
end, we identify the period $2$ transfer matrix
\beq {\cal A} = \left( \begin{array}{cc}
-\frac{\mu}{\De + w} & \frac{\De - w}{\De + w} \\
1 & 0 \end{array} \right)
\left( \begin{array}{cc}
\frac{\mu}{\De + w} & \frac{\De - w}{\De + w} \\
1 & 0 \end{array} \right), \eeq and the characteristic polynomial
\beq f(z) = \det({\cal A} - I z) = z^2 + \left( \frac{2w^2-2\De^2 +
\mu^2}{ \left( w + \De\right)^2} \right) z + \left(\frac{\De -
w}{\De+w} \right)^2. \eeq Our topological invariant (for $w + \De
\neq 0$) takes the form \beq \nu ~=~ - ~\mbox{sgn} \left( f(1) f(-1)
\right) ~=~ \mbox{sgn} \left( \left| \mu \right| - 2 \left| \De
\right| \right). \label{eq:condground} \eeq In this case, the
topological nature of the phase depends on the relative magnitude of
the superconducting order parameter and the chemical potential, as
opposed to the hopping integral and the chemical potential as in the
uniform case, cf. Eq.~(\ref{eq:condfull}) and
Eq.~(\ref{eq:condground}). That the point separating the
topologically trivial and non-trivial regions derived from $\nu=0$ is
equivalent to gap closure can be gleaned from the dispersion for $s_n
= (-1)^n$, \beq \om_k^2 ~=~ 2 ~\left[ w^2 ~+~ \De^2 ~+~
\frac{\mu^2}{2} ~+~ \left(w^2 ~-~ \De^2 \right) \cos (2k) ~\pm~ \De
\mu \sin k \right]. \eeq Specifically, consistent with
Eq.~(\ref{eq:condground}), the gap closes at $k = \pm \pi /2$ for
$2|\De|=|\mu|$. Once again, as in the full vortex sector, the
isolated Majorana modes at the end of a finite chain or at the
interface of two regions can be identified using
Eq.~\ref{eq:condground} (see Fig.~\ref{fig:sectors}e). Thus we see
that the imposition of a periodic pattern can give rise to
qualitatively different physics in that here the topological nature
of the system depends on the magnitude of $\De$.

\subsection{Higher period sectors}

The consideration of periodic patterns in the site-polarity $s_n$'s
provides a zoo of configurations for realizing topologically
non-trivial phases and interface Majorana modes. We can once again
invoke the topological invariant $\nu$ for any periodic structure $P$
characterized by the transfer matrix ${\cal A}_P$ of
Eq.~(\ref{eq:AP}). In general, we have \bea \nu_P &=& - \prod_{m =
\pm} ~\mbox{sgn}
\left( \det {\cal A}_P + 1 + m \mbox{Tr} \mathcal{A}_P \right) \non \\
&=& \mbox{sgn} \left( \left| \mbox{Tr} \mathcal{A}_P \right| - \left|
\det {\cal A}_P + 1 \right| \right). \label{eq:general} \eea It
follows from these expressions that any two chains which are cyclic
permutations of each other must have the same topology, as we would
expect physically. It is also true that the transformation $s_n
\rightarrow -s_n$ for all $n$ will not change the topology of the
chain. Although giving a full analytic expression is not practical,
it may be noted that the detailed features of a pattern of the site
polarities $s_n$ that control its topology enter
Eq.~(\ref{eq:general}) through the quantity $\mbox{Tr}
\mathcal{A}_P$. Some insight can therefore be gained by an expansion
in certain limits. For example, if $|\mu| < 2|w \pm \De|$, a useful
expansion of $\mbox{Tr} \mathcal{A}_P$ in powers of $\mu^2/(w^2 -
\De^2)$ is (for $p$ even) \beq \hspace*{-2cm} \mbox{Tr} \mathcal{A}_P
= \left(\frac{\De-w}{\De+w}\right)^{p/2} \left[ 2 + \frac{\mu^2}{2
(w^2-\De^2)} \left( \sum_{m=1}^{p/2} s_{2m} \right) \left(
\sum_{n=1}^{p/2} s_{2n-1} \right) + \mathcal{O} \left( \mu^4 \right)
\right]. \eeq Thus, in this case the quantity $\left(\sum_{m=1}^{p/2}
s_{2m} \right) \left(\sum_{n=1}^{p/2} s_{2n-1} \right)$ represents
the most relevant feature of $P$ in determining its topology. As a
specific example, taking ${\cal A}_P = A_1 A_2 A_3 A_4$ for a generic
period 4 pattern $(s_1,s_2,s_3,s_4)$, we get an expression for $\nu$
which is a function of $(s_1 + s_3)(s_2 + s_4)$, $s_1 s_2 s_3 s_4$,
and the couplings constants $w$, $\De$, and $\mu$. Thus from the
standpoint of topologically non-trivial phases, there are two
distinct period 4 patterns, namely $+ + + - $ and $+ + - - $. For
instance, the application of equation (\ref{eq:general}) to the
pattern $P = + + - - $ yields \beq \nu_P ~=~ \mbox{sgn} \left(\mu^2 -
4 | w \De | \right), \eeq as illustrated in Fig.~\ref{fig:sectors}f.
Similar constraints are also obtained for the other patterns,
demonstrating that patterns of higher periodicity can have phase
boundaries involving more intricate dependencies on the values of
$\mu$, $w$, and $\De$. Thus, we have shown that the topology and
existence of isolated Majorana modes can be controlled by
configurations of vortices in the Kitaev spin ladder.

\section{Dynamics}
\label{sec:dynamics}

Having charted out the topologically non-trivial phases and
associated Majorana physics in the Kitaev ladder, we move on to
dynamically tuning between topologically trivial and non-trivial regions.
We briefly visit this issue in the context of manipulating isolated
Majorana modes and then present a comprehensive study on dynamic
quenching between these regions.

\subsection{Majorana manipulation}

From the point of view of topological quantum computation, a system hosting
zero energy Majorana modes can in principle store
non-local quantum information. In 2D systems, gate operations on the
degenerate subspace can be achieved by the dynamic interchange of such modes,
exploiting the non-Abelian nature of such exchanges.
Recent proposals have explored schemes for transporting isolated
Majorana fermions in 1D $p$-wave superconductor wires and also braiding
them in a network of quantum wires including $T$-junctions \cite{sau,alicea}.
These protocols are based on the tunability of the bulk topology and the
controlled growth or contraction of regions of non-trivial topology.

The Kitaev ladder is well-suited to such a scheme given the many
different ways of controlling the topology of the system. Primarily,
moving an isolated Majorana mode would amount to dynamically moving
the interface between topologically trivial and non-trivial regions,
such as those shown in Fig. \ref{fig:sectors}. One method of doing so
would be to dynamically change the spatial profile of the couplings
$J_x$, $J_y$, and $J_z$ which determine the phases in the different
regions, for instance, as shown in the phase diagram of Fig.
\ref{fig:phases}. Alternatively, Majorana modes may be manipulated by
changing the vortex patterns which, as discussed in section
\ref{sec:vortex_Majorana}, determine the topology of a region. The
presence or absence of a vortex is determined by the site-polarity
$s_n$, or, equivalently, the $g$-fermion occupation number discussed
in section \ref{sec:jordan}. These occupation numbers can be
controlled by a combination of operators which do not appear in the
Kitaev ladder Hamiltonian in equation (\ref{eq:ladder}). Examples
include $J_1(\si_{2n,u}^y\si_{2n+1,u}^y-\si_{2n,l}^x\si_{2n+1,l}^x)
=2 i J_1(g^{\phantom\dagger}_n g^{\phantom\dagger}_{n+1}+g_n^\da
g^\da_{n+1} )$ and $J_2(\si_{2n,u}^y
\si_{2n+1,u}^y+\si_{2n,l}^x\si_{2n+1,l}^x) =2 i J_2(g_n^\da
g^{\phantom\dagger}_{n+1}+g^{\phantom\dagger}_n g_{n+1}^\da)$. A
possible means of reversing the site-polarities on adjacent sites
would be by the application a ``$\pi$ pulse" involving either of
these operators. Schemes for exchanging Majorana modes can be
developed using such transport and, in principle, given that the
system is a ladder, as opposed to a wire, the need for $T$-junctions
can be eliminated.


Translating between the language of qubits hosted by Majorana modes
and spin physics is a rather subtle task requiring further study. As
discussed in section \ref{sec:symbreaking}, a direct correspondence
exists between isolated Majorana operators and certain spin
operators. Care needs to be taken to ensure that the relevant degrees
of freedom are not affected by local fields, and hence, environmental
effects. Na\"{\i}vely, this seems possible given that the
Jordan-Wigner string is non-local, and at the same time, the final
read out could be more accessible than in fermionic systems by the
use of spin-sensitive probes such as magnetic cantilevers. Thus, spin
systems may afford an interesting laboratory to explore topological
phases of matter and the ideas of topological quantum computation.

\subsection{Quench dynamics}

The issue of quenching a topological 1D system through different
quantum phases is particularly germane for schemes which involve
moving bound Majorana modes localized by shifting topologically
non-trivial segments \cite{sau,alicea}. The dynamics associated with
a slow quench across a quantum critical point (QCP) has generated
considerable recent interest in its own right
\cite{kibble,zurek1,zurek2,dziarmaga1,damski,polkov1,cherng,mukherjee,deng,mondal,polkov2,dziarmaga2,dutta},
although its application to topological systems is as yet sparse. To
recapitulate some of the basic features, quenching typically involves
initializing a system in the ground state of a given Hamiltonian and
then varying a parameter of the Hamiltonian at a finite rate $1/\tau$
so as to take it across a QCP. The prominent aspect of such a quench
is that one ends up with a finite density of defects, or a residual
energy (defined below), which scales as a power of $1/\tau$. In $d$
dimensions, the defect density or residual energy obeys the
Kibble-Zurek scaling form $1/\tau^{d\nu/(z\nu +1)}$, where $\nu$ and
$z$ are respectively the correlation length and dynamical critical
exponents of the QCP
\cite{kibble,zurek1,zurek2,dziarmaga1,damski,polkov1}. These
exponents are defined by the divergence of the correlation length as
$|\lambda - \lambda_c|^{-\nu}$ and the relaxation time as $|\lambda -
\lambda_c|^{-z\nu}$, respectively, as a parameter $\lambda$ in the
Hamiltonian approaches a critical value $\lambda_c$.

Here, we consider quenching for the case extensively discussed in
previous sections, namely, the full vortex sector ($s_n=1$) in the
Kitaev ladder which maps directly onto the 1D $p$-wave
superconductor. In this case, in terms of the Majorana fermions, in
the momentum basis, theHamiltonian (\ref{eq:diracladder}) is of the
form \beq H(k) ~=~ \sum_k ~\vec{f}_k^\da \left( \begin{array}{cc}
2 w \cos k + \mu & 2 \De \sin k \\
2 \De \sin k & - 2 w \cos k-\mu \end{array} \right) \vec{f}_k,
\label{eq:Hk} \eeq where $\vec{f}_k^\dagger$ denotes the momentum
vector $(f_k^\da ~~f_{-k} )$. As detailed earlier, the system
respects the dispersion $\om_k ~=~ \pm\sqrt{(2w \cos k + \mu)^2+
4\De^2 \sin^2 k}$ and the resultant phases are as shown in Fig.
\ref{fig:phases}. We study two kinds of quenches: (A) between the two
possible topologically non-trivial phases $I$ and $II$ in Fig.
\ref{fig:phases}, and (B) between a topologically non-trivial and
topologically trivial phase, such as between phases $I$ and $III$. We
assume a linear quench in which one of the parameters of the
Hamiltonian, to be specified below, varies in time as $\al(t)=w_0
t/\tau$, so as to take the system through the relevant QCP. We
quantify the effect of the quench using the residual energy per site,
$E_r$, attained at the final time, defined as \beq E_r ~=~ \lim_{t,N
\to \infty} ~ \frac{\la H(t) \ra_f - E_0}{N|\al(t)|}, \label{eq:er1}
\eeq Here, $N$ denotes the number of sites, $\la H \ra_f$ denotes the
expectation value of the Hamiltonian in the final state reached, and
$E_0$ is the ground state energy at $t \to \infty$. We will study the
dependence of $E_r$ on $\tau$ for $\tau \gg 1$; we expect this to
vanish in the adiabatic limit $\tau \to \infty$.

For both kinds of quenches, as shown in the dispersions in Fig.
\ref{fig:phases}, the system consists of two bands which are
symmetric about the Fermi energy. In the ground state, the band below
the Fermi energy is occupied and the one above is unoccupied. Well
within one of the phases, the system is gapped. The quench takes the
system through a gapless point at a QCP (as described below) and then
into another gapped phase. For any given set of $k$-modes in
Eq.~(\ref{eq:Hk}), the quench couples the two states in the subsystem
respecting $f_k^\da f^{\phantom\dagger}_k+f_{-k}^\da
f^{\phantom\dagger}_{-k}=1$. Thus, the relevant sector contributing
to the quench dynamics comprises of the states having occupation
numbers $|n_k,n_{-k}\rangle$ equal to $|1,0\rangle$ and
$|0,1\rangle$. In this basis, the quenches for cases (A) and (B) take
the form \bea {\cal H}_{Ak}(t) &=& (2w_0\cos k + \mu_0)~\tau^z ~+~
\al(t) (2\sin k) ~\tau^x,
\non \\
{\cal H}_{Bk}(t) &=& (2w_0\cos k + \al(t))~\tau^z ~+~ \De_0 (2\sin k) ~\tau^x,
\label{eq:Hquench} \eea
where $w_0$, $\mu_0$ and $\De_0$ correspond to some fixed parameter
values, and $\tau^a$ denote the Pauli matrices. Thus, case (A) involves
tuning through $\De=0$ for fixed $|\mu_0 /(2w_0)| < 1$ and case (B) through
$\mu = 2w_0$ for fixed $\De_0/w_0$. These take the system through a gapless
point which lies at wave vectors $k_c= \pm \cos^{-1} [-\mu_0/ (2w_0)]$ in
case (A) and $k_c=0$ and $\pi$ in case (B). In both cases, the critical
exponents are given by $z=\nu=1$, since the gap is (i) linear in $|k-k_c|$
at the QCP $\al=0$, and (ii) linear in $|\al|$ when $\al$ is close to zero,
for $k=k_c$.


For the linear quench $\al(t)=w_0 t/\tau$, the dynamics in both cases
maps exactly to the well-known Landau-Zener problem, as with a host
of other quenches
\cite{zurek1,zurek2,dziarmaga1,damski,polkov1,cherng,mukherjee,deng,lz}.
In particular, for case (B), the probability $p_k$ of ending in an
excited state at time $t \gg \tau$ and the net residual energy, for
which each sub-system contributes $4p_k$, are given by \beq p_k ~=~
\exp [- \pi (2 \De_0 \sin k)^2 \tau /w_0], ~~~{\rm and}~~~ E_r ~=~
\int_0^{\pi} ~ \frac{dk}{2\pi} ~4 p_k. \eeq The probability $p_k$ is
largest for the low-energy modes near $k_c = 0$ and $\pi$ where the
gap vanishes at the QCP. In the limit $\tau \gg w_0 / \Delta_0^2$,
the residual energy is dominated by these modes; expanding around
$k_c$ gives a Gaussian integral and the power-law form $E_r \sim
1/\tau^{1/2}$ which is consistent with the Kibble-Zurek power
$d\nu/(z\nu +1)$ given that $d=z=\nu =1$. The same power-law is
obtained for case (A), although the expression for the excitation
probability $p_k$ looks a bit different from case (B).

In principle, quenching can be studied in the other vortex sectors
discussed above. For most sectors, the problem becomes analytically
intractable even for a case in which $\De = 0$ and only $\mu$ is varied
in time linearly. As shown by two of us in previous work \cite{senvish},
an interesting situation arises when a pattern of $s_n$ is endowed
with a higher symmetry in that some matrix elements of the
Hamiltonian in equation (\ref{eq:kham}) are equal to zero.
For instance, in the sector in which the signs of $s_n$ form the
period 4 pattern ($++--$), we find that the low-energy modes at $k=0$
and $\pi$ (which dominate the generation of residual energy) are
coupled via a single intermediate high-energy state lying at
$k=-\pi/2$ and $\pi/2$. Using second order perturbation theory to
eliminate the high-energy states, we find that the effective
time-dependent Hamiltonian governing the low-energy states is given by
\beq H_{eff} ~=~ \left( \begin{array}{cc}
c t^2/\tau^2 & 2 k \\
2 k & - c t^2/\tau^2 \end{array} \right), \eeq where the variable $k$
has been redefined to lie close to 0, and $c$ is a constant. Using a
scaling analysis, one can then show that the residual energy goes as
$E_r\sim 1/\tau^{2/3}$ \cite{senvish}. Further work would involve
studying such anomalous scaling for quenches into topologically
non-trivial phases involving $\Delta \neq 0$.

More generally, if the most relevant states participating in the quench are
not directly coupled, but instead are coupled via $q$ intermediate steps, it
can be argued that the residual energy, for a linear quench, scales
as $E_r\sim 1/\tau^{(q+1)/(q+2)}$. Thus, the quench dynamics in these
sectors is also controlled by the QCP, resulting in a power-law
scaling of the relevant quantities with respect to the quench rate.

\section{Applications to the 1D $p$-wave superconductor}
\label{sec:appl_1Dpwave}

Given that most of our analysis of the Kitaev ladder involves mapping
to the 1D $p$-wave superconductor, several of our results can be
directly applied to the latter system and related ones.

(i) {\it Periodic potentials and topologically non-trivial phases:-} One
of our main results is that vortex patterns in the Kitaev model can alter
the topology of the system. The cases that we considered involved
periodic patterns in the signs $s_n$.
The primary effect of a periodic pattern is to introduce more
bands in momentum space, an effect which can equally well be achieved
by the presence of a superlattice or a periodic externally applied potential.
Thus, we predict that in the $p$-wave superconducting system, an alternate
means of changing the topological nature of a wire and the associated
Majorana end modes is the application a periodic potential. We have developed
a straightforward technique for identifying these features even for complex
dispersions based on a topological invariant directly derived from equations
of motion. Our analysis provides detailed results for the topologically
non-trivial phases
and isolated Majorana modes in the special case of the external potential
controlling only the sign of the local chemical potential on each site. Our
results show that such a potential can alter the physics in qualitative ways,
for instance, in the case of an alternating
site potential, enabling the superconducting gap function to act as a
tuning parameter for achieving topologically non-trivial phases.

(ii) {\it Quench dynamics:-} The quench analysis above provides a word
of caution for schemes which entail dynamically changing the
topological nature of wire segments. In the thermodynamic limit, our
analysis shows that no matter how slowly the system is tuned from one
phase to another, the presence of the QCP results in defects whose
density has a $1/ \tau^{1/2}$ power-law dependence on the tuning rate $1/
\tau$. Thus, except in the extreme adiabatic limit, the quench is
bound to generate defects which could cause quantum decoherence. Even in
the case of changing a finite segment from a topologically non-trivial
to a topologically trivial region, it should be kept in mind
that the eigenstates of the initial system span its entire length and
a local change, particularly involving a gapless point, could affect
the entire system and produce a finite density of defects. In fact,
the route to equilibrium upon quenching a finite region is in itself
an interesting problem which could place stringent constraints on
proposed dynamical schemes.

\section{Conclusions}
\label{sec:concl}

In conclusion, we have introduced and analyzed a ladder version of
Kitaev's honeycomb model. By employing a mapping to a 1D $p$-wave
superconducting system, we have borrowed from the insights provided
by this system in terms of topological features. We find an intimate
connection between spontaneous symmetry breaking and isolated
Majorana modes; this connection provides us with an alternative view
of topological order and offers an attractive avenue for the
investigation of a wide class of spin systems.

We have performed a detailed study of the ladder system in the presence of
vortex arrays and shown that the presence of vortices can dramatically alter
the topological properties of the system. We have identified various
conditions and configurations of vortices that
result in the isolation of Majorana modes, states which are currently being
considered as prime candidates for the building blocks in topological
quantum computation. We have instigated the study of quench dynamics
across quantum critical points in these topological systems, discussing the
power-law scaling of the residual energy density; we believe that this avenue
unveils important physics from the perspective of non-equilibrium quantum
critical phenomena as well as in realistic treatments of the dynamic
manipulation of topological entities in quantum computational schemes.

Our findings are germane to various physical systems that are
currently under study for their topological properties. Several of
our results are directly applicable to quantum wires which, by various
methods, can effectively host proximity induced superconducting order
\cite{sau,sau2,oreg,alicea,neupert,mao,duck,potter,chung,hosur,ganga}.
In the past years, several proposals
have focused on realizing the Kitaev honeycomb system in cold atoms trapped
in an optical lattice \cite{zhang,han}; our results can be immediately
translated to these systems. Furthermore, our studies show that it is
possible to build a dictionary between topological aspects of non-interacting
fermionic systems and ordering in spin systems; an extensive dictionary would
enable us to apply insights found in one physical system to the other.

\section*{Acknowledgments}

This work is supported by the National Science Foundation under the
grant DMR 0644022-CAR (W.D. and S.V.) and DST, India under Project
No. SR/S2/CMP-27/2006 (D.S.). We would like to thank T. Hughes for
detailed and illuminating discussions and D. Ferguson and J.
Slingerland for their perceptive comments.

\end{document}